\documentclass[10pt]{iopart}

\usepackage{graphicx}
\usepackage{amssymb,amsfonts,dsfont}
\usepackage{customShortcuts}
\usepackage{color}
\usepackage{array}
\usepackage{upgreek}
\usepackage{mathrsfs}
\usepackage[normalem]{ulem}

\expandafter\let\csname equation*\endcsname=\relax
\expandafter\let\csname endequation*\endcsname=\relax
\usepackage{amsmath}

\newcommand{\Vrms}{V^0_\mrm{rms}}
\newcommand{\wrs}{\w_\mrm r}

\newcommand{\edt}{\eps_\mrm d}

\newcommand{\csp}{\chi_\mrm s}
\newcommand{\crs}{\chi_\mrm r}
\newcommand{\kL}{\ket{\mrm L}}
\newcommand{\kR}{\ket{\mrm R}}
\newcommand{\bL}{\bra{\mrm L}}
\newcommand{\bR}{\bra{\mrm R}}
\newcommand{\pL}{\proj{\mrm L}}
\newcommand{\pR}{\proj{\mrm R}}

\begin{document}

\title{Coupling a single electron spin to a microwave resonator: Controlling transverse and longitudinal couplings}
\author{F\'elix Beaudoin$^1$, Dany Lachance-Quirion$^{2,3}$, W. A. Coish$^{1,4}$, Michel Pioro-Ladri\`ere$^{2,3,4}$}
\address{$^1$ Department of Physics, McGill University, Montr\'eal, Qu\'ebec H3A 2T8, Canada\\
	$^2$ 	D\'epartement de Physique, Universit\'e de Sherbrooke, Sherbrooke, Qu\'ebec J1K 2R1, Canada\\
	$^3$ Institut Quantique, Universit\'e de Sherbrooke, Sherbrooke, Qu\'ebec J1K 2R1, Canada\\
	$^4$ Quantum Information Science Program, Canadian Institute for Advanced Research, Toronto, Ontario M5G 1Z8, Canada
	}

\date{\today}

\begin{abstract}

Microwave-frequency superconducting resonators are ideally suited to perform dispersive qubit readout, to mediate two-qubit gates, and to shuttle states between distant quantum systems. A prerequisite for these applications is a strong qubit-resonator coupling.  Strong coupling between an electron-spin qubit and a microwave resonator can be achieved by correlating spin- and orbital degrees of freedom. This correlation can be achieved through the Zeeman coupling of a single electron in a double quantum dot to a spatially inhomogeneous magnetic field generated by a nearby nanomagnet. In this paper, we consider such a device and estimate spin-resonator couplings of order $\sim 1$~MHz with realistic parameters. Further, through realistic simulations, we show that precise placement of the double dot relative to the nanomagnet allows to select between a purely longitudinal coupling (commuting with the bare spin Hamiltonian) and a purely transverse (spin non-conserving) coupling.  Additionally, we suggest methods to mitigate dephasing and relaxation channels that are introduced in this coupling scheme. This analysis gives a clear route toward the realization of coherent state transfer between a microwave resonator and a single electron spin in a GaAs double quantum dot with a fidelity above $90$\%. Improved dynamical decoupling sequences, low-noise environments, and longer-lived microwave cavity modes may lead to substantially higher fidelities in the near future.

\end{abstract}

\ioptwocol

\section{Introduction}

After nearly two decades of development, a wide range of quantum-information-processing devices with complementary capabilities have emerged~\cite{wallquist2009hybrid,xiang2013hybrid}. For example, qubits based on superconducting elements now show exceptionally high-fidelity quantum logic operations, regularly exceeding the fidelity of $\sim 99\%$ required for fault-tolerant quantum computation~\cite{o2015qubit,corcoles2013process,sheldon2016procedure}.  This level of control has now enabled multiple rounds of error correction to protect against bit flips~\cite{kelly2015state}, allowing for the protection of classical states.  Physical-qubit memory times for quantum states in superconducting devices are, however, typically limited to $\lesssim 100\,\mu\mathrm{s}$. In contrast, spin qubits in semiconductors show extremely long phase memory times without error correction, reaching $870\,\mu\mathrm{s}$ for electron spins in GaAs quantum dots~\cite{malinowski2016notch}, $28\,\mathrm{ms}$ for electron spins in silicon quantum dots~\cite{veldhorst2015two}, exceeding $0.5\,\mathrm{s}$ for electron spins at phosphorus donor impurities in a silicon device~\cite{muhonen2014storing}, and exceeding $30\,\mathrm{s}$ for nuclear spins at phosphorus donors in the same silicon device~\cite{muhonen2014storing}.  Experiments showing a coherence time of $3$ \emph{hours} for nuclear spins in isotopically enriched silicon (in bulk) \cite{saeedi2013room} suggest that current spin-qubit devices are far from reaching any fundamental limit in memory time. Transferring quantum information between superconducting and spin qubits could provide obvious advantages, maximizing both control fidelity and memory time in a future quantum processor.

A natural intermediary between superconducting qubits and spin qubits is a microwave resonator \cite{wu2010storage,amsuess2011cavity,kubo2011hybrid,kubo2012storage,probst2013anisotropic,kloeffel2013circuit}, which supports quantized electromagnetic modes that leak out of the resonator with rate $\kappa$. A prerequisite to using such resonators for high-fidelity information transfer is that the qubit-resonator system be in the strong-coupling regime, $g>\kappa$, where $g$ is the qubit-resonator coupling. Several schemes have been proposed to achieve strong coupling between a microwave resonator and qubits encoded in one, \cite{trif2008spin,cottet2010spin,hu2012strong} two, \cite{burkard2006ultra,jin2012strong} or three \cite{russ2015long,srinivasa2016entangling} electron spins in quantum dots, or encoded in the electron and nuclear spins of phosphorus donor impurities in silicon \cite{tosi2015silicon}. Readout of a spin qubit through a combination of spin-to-charge conversion and charge sensing with a microwave resonator has been demonstrated experimentally~\cite{petersson2012circuit}.  Very recently, the strong-coupling regime has been reached experimentally for a single spin in a carbon-nanotube double quantum dot coupled to a microwave resonator, although coherent transfer of quantum information between these two systems has not yet been demonstrated~\cite{viennot2015coherent}.

The direct magnetic coupling of a single spin to the electromagnetic field of a microwave resonator is weak, $g\lesssim 100\,\mathrm{Hz}$~\cite{schoelkopf2008wiring,imamoglu2009cavity}.  To approach strong coupling, the proposals and experiments listed above therefore correlate the spin and charge degrees of freedom through spin-orbit coupling,~\cite{trif2008spin,petersson2012circuit} exchange coupling between spins in quantum dots,~\cite{burkard2006ultra,jin2012strong,russ2015long,srinivasa2016entangling} or through spatially varying magnetic or exchange fields from a nearby ferromagnet~\cite{hu2012strong,cottet2010spin,viennot2015coherent}. Large exchange-field or magnetic-field gradients $\sim1\;\mrm T/\upmu$m have been generated near quantum dots in carbon nanotubes~\cite{viennot2015coherent}, GaAs quantum dots~\cite{pioro2008electrically} and Si/SiGe quantum dots~\cite{kawakami2014electrical}.

In this paper, we consider a single spin in a double quantum dot exposed to an inhomogeneous magnetic field.  In this configuration, the spin-resonator coupling Hamiltonian can be divided into two distinct terms: one that commutes with the Zeeman Hamiltonian (longitudinal) and one that does not (transverse). While the transverse coupling may be used for quantum state transfer between the qubit and the resonator, the longitudinal coupling could be used in a novel two-qubit gate~\cite{jin2012strong,billangeon2015circuit,royer2016fast} and in qubit readout~\cite{didier2015fast} schemes introduced recently for superconducting qubits. We evaluate the strengths of these two couplings in a realistic GaAs/AlGaAs device using simulations of the magnetic-field configuration due to a nanomagnet. Proper placement of the nanomagnet above the double quantum dot allows for either a purely transverse or purely longitudinal coupling. Finally, we investigate the additional dephasing and relaxation channels that arise due to an admixture of spin- and orbital degrees-of-freedom in the presence of the nanomagnet~\cite{tokura2006coherent}. We show how a combination of dynamical decoupling and careful optimization of parameters can lead to experiments demonstrating a coherent state transfer between a resonator and a single spin in a GaAs device with a fidelity above 90\%. Our analysis of error sources also indicates that an even higher fidelity should be achievable in silicon.

\section{Spin-resonator couplings	\label{secInteractions}}

In this section, we describe the physical setup under consideration and derive the effective Hamiltonian for a single electron spin in a double quantum dot coupled to a microwave resonator.  In addition to the transverse (spin-non-conserving) coupling that has been widely studied in the literature, we consider a longitudinal (spin-conserving) spin-resonator coupling and comment on some applications of this type of coupling.

\begin{figure}
  \includegraphics[width=0.2\textwidth]{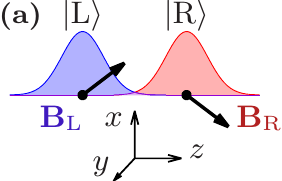}
  \hspace{0.06\textwidth}
  \raisebox{5.75mm}{\includegraphics[width=0.19\textwidth]{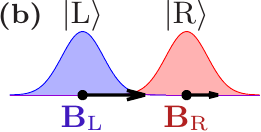}}
 
  \raisebox{20mm}{\textbf{(c)}}
  \hspace{0.025\textwidth}
  \includegraphics[width=0.35\textwidth]{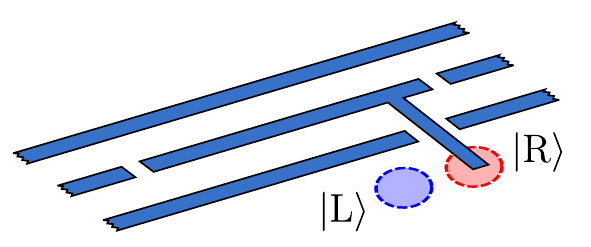}
  
 \begin{center}
  \includegraphics[width=0.48\textwidth]{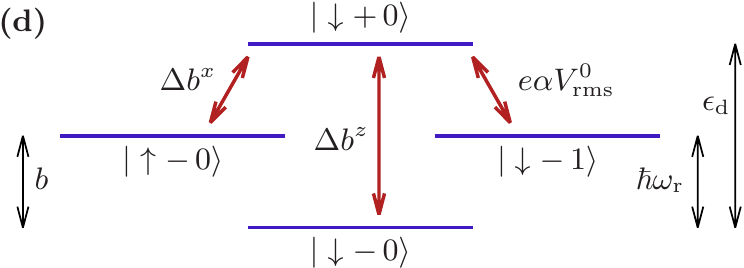}
 \end{center}
 \caption{Scheme to achieve strong coupling of a single electron spin to a coplanar waveguide resonator. (a), (b) Spin-charge coupling is achieved in a double quantum dot with left ($\ket {\mrm L}$) and right ($\ket {\mrm R}$) orbital states by means of an inhomogeneous magnetic field. $\mvec B_\mrm{L/R}$ is the magnetic field associated with the orbital state $\ket{\mrm{L/R}}$. Inhomogeneity in the magnetic field may be (a) transverse, or (b) longitudinal with respect to the quantization axis of the electron spin, $z$. (c) Charge-resonator coupling is achieved through capacitive coupling to the right (or left) dot only. (d) Energy level structure of the system in the basis formed by the bare eigenstates of the electron spin ($\upa$, $\dwna$), double quantum dot ($\pm$), and resonator ($n=0,1,...$), in the absence of spin-charge and charge-resonator coupling. Introducing spin-charge ($\propto \D b^x$, $\D b^z$) and charge-resonator ($\propto e\al \Vrms$) couplings generates virtual transitions indicated by the red arrows, where $\D b^x=b_\mrm L^x-b_\mrm R^x$ and $\D b^z=b_\mrm L^z-b_\mrm R^z$, with $b^{x,z}_\mrm{L(R)}$ the $x$ and $z$ components of the left (right) Zeeman field, $e$ is the elementary charge, $\Vrms$ is the zero-point voltage of the resonator and $\al$ is the lever arm of the dot-resonator coupling. In addition, $\eps_\mrm d$, $b$, and $\hbar\w_\mrm r$ are the double-dot, spin, and resonator energy splittings, respectively.
 \label{figDevice}}
\end{figure}

For concreteness, we consider the device illustrated in Fig.~\ref{figDevice}(c) consisting of a double quantum dot capacitively coupled to a coplanar-waveguide resonator. In addition,  we consider an inhomogeneous magnetic field [Figs.~\ref{figDevice}(a,b)], and the associated Zeeman coupling.  The electron then experiences a different magnetic field $\mvec B_l\equiv\int d\mvec r |\psi_l(\mvec r)|^2\mvec B(\mvec r)$ in the lowest orbital state of dot $l=\mrm L (\mrm R)$, for the left (right) dot with the associated envelope function $\left<\mvec r\right|\left.l\right>=\psi_l(\mvec r)$. Here, we take the direction of the average magnetic field $\mvec B\equiv(\mvec B_\mrm L+\mvec B_\mrm R)/2$ to define the $z$ axis through the Zeeman field, $g^*\mu_B\mvec B=b\mvec{\hat z}$ (with $g$-factor $g^*$, Bohr magneton $\mu_\mathrm{B}$, and Zeeman splitting $b>0$). Since an electron in the left or right orbital experiences a distinct magnetic field coupling to its spin, the inhomogeneous magnetic field induces a coupling between spin and orbital degrees-of-freedom proportional to the difference field, $\D\mvec B\equiv\mvec B_\mrm L-\mvec B_\mrm R$~\cite{pioro2008electrically,yoneda2014fast}. In addition, the right dot is taken to be sensitive to the zero-point voltage $\Vrms$ of a coplanar-waveguide resonator through a capacitive finger [Fig.~\ref{figDevice}(c)] with lever arm $\al$, which couples microwave photons in the resonator to the orbital degree-of-freedom of the electron~\cite{childress2004mesoscopic,frey2012dipole}. Introducing the double-dot detuning $\varepsilon$, tunnel splitting $\Omega$, and resonator frequency $\wrs$, we then model this device with the Hamiltonian
\begin{align}
 H=&\;{\textstyle \frac12} \left(\veps\tau_z+\W\tau_x\right)+\hbar\wrs a^\dag a+{\textstyle\frac12}b\s_z\notag+{\textstyle\frac14}\mvec{\D b}\cdot\mvec \s\,\tau_z\\
 &+e \al \Vrms(a^\dag+a)(1-\tau_z)/2.	\label{eqnHlocal}
\end{align}
Here, $a$ annihilates a microwave photon in the resonator, $\mvec \s=(\sigma_x,\sigma_y,\sigma_z)$ is the vector of Pauli operators describing the electron spin and we have also introduced Pauli operators associated with the orbital states: $\tau_x=\kL\bR+\kR\bL$ and $\tau_z=\pL-\pR$ [note, in particular, that $(1-\tau_z)/2 = \pR$ in the second line of Eq.~\eqref{eqnHlocal}].  The difference field leads to the term $\mvec{\D b}\equiv g^\ast\mu_B\mvec{\D B}=g^\ast\mu_\mrm B(\mvec B_\mrm L - \mvec B_\mrm R)$. Without loss of generality, in what follows we choose a coordinate system so that $\mvec{\D b}=\D b^x\mvec{\hat x}+\D b^z\mvec{\hat z}$ lies in the $x$-$z$ plane. 

The orbital part of Eq.~\eqref{eqnHlocal} can be diagonalized by transforming to the basis of double-dot eigenstates $\ket{+}=\cos\frac{\theta}{2}\kL+\sin\frac{\theta}{2}\kR$ and $\ket-=-\sin\frac{\theta}{2}\kL+\cos\frac{\theta}{2}\kR$, with associated eigenenergies $\pm\epsilon_\mathrm{d}/2=\pm\sqrt{\varepsilon^2+\Omega^2}/2$, and where $\tan\q=\Omega/\varepsilon$ (see \ref{appSW}).  We then rewrite the transformed total Hamiltonian ($\tilde{H}=R_y^\dagger(\theta)HR_y(\theta)$, $R_y(\theta)=e^{-i\theta\tau_y/2}$) as $\tilde{H}=\tilde{H}_0+\tilde{V}$, separating $\tilde{H}$ into contributions that are purely diagonal ($\tilde{H}_0$) and purely off-diagonal ($\tilde{V}$) in the basis of states $\ket{s\,d\,n}$, where $\sigma_z$-eigenstates are labeled by $s\in\left\{\uparrow,\downarrow\right\}$, double-dot orbital states are labeled by $d\in\left\{+,-\right\}$, and $n\in\left\{0,1,2,\ldots\right\}$ gives the number of microwave photons in the resonator.  The off-diagonal contribution $\tilde{V}$ contains spin-charge ($\propto \D b^x,\,\D b^z$) and charge-resonator couplings ($\propto e\al\Vrms$), and thus couples the spin to the resonator through virtual transitions to the orbital excited state $\ket+$ of the double quantum dot, as shown in Fig.~\ref{figDevice}(d) (red arrows). The term proportional to $\D b^x\ne 0$ [magnetic-field configuration in Fig.~\ref{figDevice}(a)] creates or destroys an orbital excitation while flipping the spin. This leads to a transverse (spin-non-conserving) coupling to the resonator.  In contrast, the term $\propto\D b^z \ne 0$ [magnetic field configuration in Fig.~\ref{figDevice}(b)] creates or destroys an orbital excitation while preserving $\s_z$, leading to a longitudinal (spin-conserving) coupling. 

To derive a low-energy effective Hamiltonian, we find a Schrieffer-Wolff transformation $e^S \tilde{H}e^{-S}$ that eliminates the off-diagonal coupling $\tilde{V}$ at leading order. Taking $\epsilon_\mrm d>b,\hbar\wrs$, we then project onto the (dressed) orbital ground-state manifold of the double quantum dot, $\{\ket{s-n}'\}=\{e^{-S}\ket{s-n}\}$.  Neglecting a constant shift and counter-rotating terms in a rotating-wave approximation, we arrive at the following projected effective spin-resonator Hamiltonian, valid to second order in $\tilde V$ (see \ref{appSW} for details):
\begin{eqnarray}
H' & = H_0'+V',\label{eqnHprime}\\
H_0'&=\frac{b'}2\s_z+\hbar\wrs' a^\dag a,\label{eqnH0prime}\\
V'&=\hbar g_x(a^\dag\s_-+a\s_+)+\hbar g_z(a+a^\dag)\s_z,\label{eqnVprime}	
\end{eqnarray}
where $b'=b+\csp-\tilde\chi_\mrm s$ and $\hbar\wrs'=\hbar\wrs-\crs$ with energy shifts $\chi_\mrm s$, $\tilde\chi_\mrm s$, and $\chi_\mrm r$ given explicitly in \ref{appSW}.  These shifts can be incorporated directly into the spin-resonator dynamics, but for simplicity we will take $b'\simeq b$ and $\omega_\mathrm{r}'\simeq\omega_\mathrm{r}$ in our explicit calculations and estimates that follow.  

In Eq.~\eqref{eqnVprime}, we have dropped counter-rotating terms $\propto \Delta b^x\left(a^\dagger\sigma_+ + a\sigma_-\right)$ and squeezing terms $\propto e\al\Vrms(a^2+a^{\dag 2})$ (sufficient conditions for this are $b,\,\hbar\wrs\gg \D b^x,e\al\Vrms$, and $|b-\hbar\wrs|\lesssim \hbar g_x$), but we retain longitudinal-coupling terms $\propto \Delta b^z\sigma_z(a+a^\dagger)$, which can become resonant if the parameter $g_z$ is modulated (see below). The transverse- and longitudinal-coupling parameters are given by
\begin{align}
 \hbar g_x\!=&\D b^x\frac{e\al \Vrms\W^2}{8\epsilon_\mathrm{d}}\!\left(\!\frac{1}{\epsilon_\mathrm{d}^2-b^2}\!+\!\frac{1}{\epsilon_\mathrm{d}^2-\hbar^2\wrs^2}\!\right)\!,	\label{eqngx}\\
 \hbar g_z\!=&\D b^z\frac{e\al \Vrms\W^2}{8\epsilon_\mathrm{d}}\!\left(\!\frac{1}{\epsilon_\mathrm{d}^2}\!+\!\frac{1}{\epsilon_\mathrm{d}^2-\hbar^2\wrs^2}\!\right)\!.	\label{eqngz}
\end{align}
In general, there are corrections to $H'$ at third and higher order in the off-diagonal coupling $\tilde{V}$ (see \ref{appSW}).  Neglecting these corrections is justified provided all off-diagonal matrix elements are small compared to the associated excitation energies, i.e.: $|\Delta b^z|\ll\epsilon_\mathrm{d}, |\Delta b^x|\ll|\epsilon_\mathrm{d}\pm b|,|e\alpha V_\mathrm{rms}^0|\ll|\epsilon_\mathrm{d}\pm\hbar\omega_\mathrm{r}|$ [see Fig.~\ref{figDevice}(d)].

The coupling parameters $g_x$ and $g_z$ can be controlled through rapid electric tuning of either the double-dot detuning $\varepsilon$ or the tunnel splitting $\Omega$ (recall $\epsilon_\mathrm{d}=\sqrt{\varepsilon^2+\Omega^2}$).  To avoid Landau-Zener transitions to the excited orbital manifold $\left\{\ket{s+n}'\right\}$ (which would invalidate the use of the projected effective Hamiltonian), $\varepsilon$ and $\Omega$ must still be tuned sufficiently slowly.  For example, for the double-dot detuning $\varepsilon$, the standard condition for an adiabatic Landau-Zener sweep is \cite{landau1932theory,zener1932non-adiabatic}:
\begin{equation}
\left|\frac{\hbar d\varepsilon}{dt}\right| < 2\pi\Omega^2.\label{eqnLZCondition}
\end{equation}
Provided Eq.~\eqref{eqnLZCondition} is satisfied, rapid manipulation of the detuning can be used to perform useful quantum operations. Indeed, the coupling $g_x$ is suppressed when taking $\ve\gg\W$, and maximized at $\ve=0$. Taking the spin and the resonator to be resonant ($\hbar\wrs=b$) and briefly tuning $\ve$ to zero (resulting in a large $g_x$) then generates a quantum state transfer between the spin and the resonator through the transverse coupling~\cite{hofheinz2008generation}. This transverse coupling is also a useful resource for two-qubit gates~\cite{blais2007quantum,majer2007coupling,chow2011simple} and readout~\cite{wallraff2004strong,reed2010high}.

To analyze the longitudinal coupling, we go to the interaction picture with respect to $H_0'$: $V'_I(t)=e^{iH_0' t/\hbar}V'e^{-iH_0' t/\hbar}$. The longitudinal-coupling term in $V'_I(t)$ ($\propto g_z$) then oscillates at angular frequency $\wrs$. This term is thus negligible under a rotating-wave approximation when $\wrs\gg g_z$ for static $g_z$. However, it is possible to make this term resonant through parametric modulation of $g_z$ at the dressed frequency of the resonator, $g_z(t)=g_z^0+\D g_z\cos(\wrs t)$. The longitudinal coupling then leads to a significant qubit-state-dependent displacement of the cavity state which, in combination with homodyne detection, produces a qubit readout. It has been shown recently that this longitudinal readout is faster than the usual dispersive readout, which relies on transverse coupling~\cite{didier2015fast}. In addition, a parametric modulation of the longitudinal coupling has recently been proposed to realize fast and high-fidelity two-qubit gates~\cite{royer2016fast}.

\section{Implementation \label{secImplementation}}

\begin{figure}
 \centering
 \includegraphics*[width = 1.00\columnwidth]{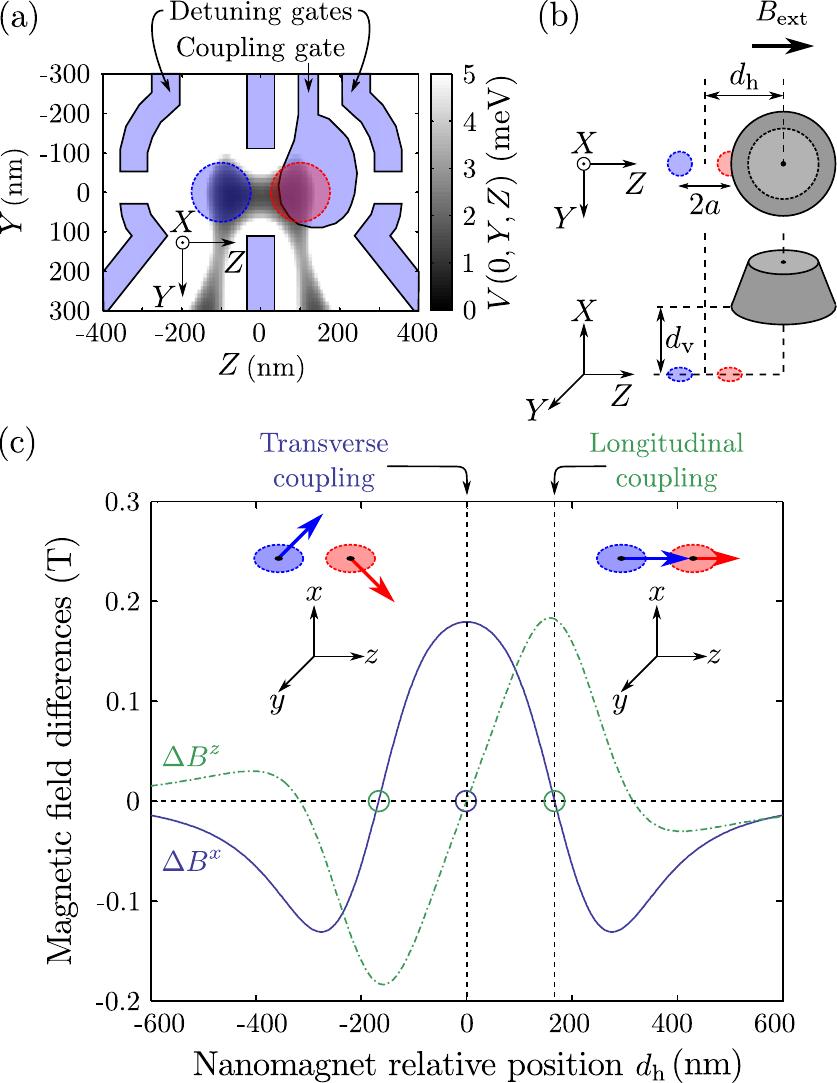}
 \caption{Spin-charge coupling with a nanomagnet. (a) Depletion gates patterned on the surface (at vertical position $X=d_\mathrm{v}=100$~nm) of a  GaAs/AlGaAs heterostructure are used to define a double quantum dot in the 2DEG ($X=0$) as depicted in the simulation of the electrostatic potential (shaded region). The coupling gate is connected to the central conductor of a superconducting coplanar waveguide resonator as in Fig.~\ref{figDevice}(c). (b) Top and side view of the nanomagnet design considered. The center of a truncated-cone shaped FeCo nanomagnet of magnetization $M=1.93$~T, base radius $220\,\mathrm{nm}$, top radius $150\,\mathrm{nm}$, and thickness $300\,\mathrm{nm}$ is placed at the position $(d_\mathrm{h},0,d_\mathrm{v})$. (c) Transverse ($\Delta B^x$, blue line) and longitudinal ($\Delta B^z$, green dot-dashed line) magnetic field differences as a function of horizontal displacement $d_\mathrm{h}$ for a quartic potential [Eq.~\eqref{eq:quartic}] with GaAs double-dot parameters $a=75$~nm, $\hbar\omega_0=1.25$~meV and $m^*=0.067m_e$ (where $m_e$ is the bare electron mass), resulting in a tunnel splitting $\Omega/h=19.1$~GHz. The external field $B_\mathrm{ext}$ is fixed to 0.675~T, giving a Zeeman splitting $b/h=3$~GHz for a $g$-factor $g^*=-0.44$ appropriate for bulk GaAs. Insets: schematic representations of the average magnetic fields in each dot for $d_\mathrm{h}=0$ (transverse coupling) and $d_\mathrm{h}=168\,\mathrm{nm}$ (longitudinal coupling).}
 \label{device_and_simulations}
\end{figure}

In this section, we perform detailed modeling for a specific device that could realize the transverse and longitudinal spin-resonator couplings derived in Sec.~\ref{secInteractions}. We consider a GaAs/AlGaAs heterostructure with a two-dimensional electron gas (2DEG) at the position $X=0$. As shown in Fig.~\ref{device_and_simulations}(a), depletion gates at the surface ($X=d_\mathrm{h}=100$~nm) are used to electrostatically define a double quantum dot in the 2DEG. The origin of the coordinate system $(X,Y,Z)$ lies at the centroid of the mirror-symmetric double dot (when $\varepsilon =0$), in the plane of the 2DEG. The system of coordinates $(X,Y,Z)$ used in this section to describe the device geometry does not necessarily coincide with the system $(x,y,z)$ used in Sec.~\ref{secInteractions} to describe the inhomogeneous magnetic field. To provide the spin-charge coupling term $\propto\Delta\mathbf b\cdot\mvec\s\tau_z$ given in Eq.~\eqref{eqnHlocal}, the double dot is exposed to the stray magnetic field of a nanometer-scale ferromagnet deposited on the surface of the heterostructure [Fig.~\ref{device_and_simulations}(b)]. 

Figure~\ref{device_and_simulations}(a) shows the electrostatic potential $V(\mvec\rho)$ in the 2DEG, where $\mvec\rho=(0,Y,Z)$ is the vector in the 2DEG plane, simulated using \textit{nextnano}~\cite{birner2007nextnano} for a given set of gate voltages. The potential $V(\mvec\rho)$ in the double dot is well-approximated by a quadratic potential along the $Y$-direction and a tilted quartic potential along the $Z$-direction~\cite{jin2012strong,burkard1999coupled}
 \begin{align}
  V(\mvec\rho)=\frac{m^\ast\omega_0^2}{2}\left[\frac{1}{4a^2}\left(Z^2-a^2\right)^2+Y^2\right]-\ve\frac{Z}{a}.
  \label{eq:quartic}
 \end{align}
We will therefore use this analytic form in simulations and estimates that follow.  In Eq.~\eqref{eq:quartic}, we have introduced the inter-dot distance $2a$, the confinement energy $\hbar\omega_0$, and the effective mass $m^\ast$. The double-dot detuning, $\ve$, was introduced in Sec.~\ref{secInteractions}. To accurately determine both the effective Zeeman term and splitting $\epsilon_\mathrm{d}$ of the orbital ground-state doublet, we numerically solve the 2D Schr\"odinger equation with the potential $V(\mvec\rho)$ defined in Eq.~\eqref{eq:quartic}.  Unless otherwise specified, throughout this paper we choose parameters that are appropriate for a typical GaAs double quantum dot: $m^*=0.067\,m_e$ ($m_e$ is the bare electron mass), $g^*=-0.44$, $\hbar\omega_0=1.25\,\mathrm{m}e\mathrm{V}$.  

\subsection{Spin-charge and charge-resonator couplings	\label{secIndiv}}

To engineer a spin-orbital coupling, we consider a FeCo nanomagnet~\cite{lachance2015magnetometry} at the surface of the heterostructure and displaced horizontally by $d_\mathrm{h}$ from the center of the double quantum dot along the $Z$-direction [Fig.~\ref{device_and_simulations}(b)]. The stray magnetic field $\mvec B_\mrm M(\mvec\rho)$ created by the nanomagnet when magnetized at saturation by an external in-plane magnetic field $\mvec B_\mrm{ext}=(0,0,B_\mrm{ext})$ along the $Z$-direction is simulated with \textit{Radia}~\footnote{The Mathematica Radia package is available at http://www.esrf.eu/.}. Recent experiments have shown that such simulations are quantitatively accurate~\cite{lachance2015magnetometry,yoneda2015robust}. 

The total wavefunctions for the double-dot ground-state doublet are $\psi_\pm(\mathbf{r})=\left<\mathbf{r}\right|\left.\pm\right>=\chi_0(X)\phi_\pm(\mvec\rho)$, where $\chi_0(X)$ is the envelope function for the lowest subband of the 2DEG and $\phi_\pm(\mvec\rho)$ are the 2D envelope states found from numerically solving the Schr\"odinger equation.  For a typical 2DEG well width of $\sim 5\,\mathrm{nm}$, the magnetic field and 2D double-dot potential vary slowly (on a length scale $\sim 100\,\mathrm{nm}$) when $\chi_0(X)$ is appreciable.  To a good approximation, the field experienced by an electron in state $\psi_\pm(\mathbf{r})$ is then independent of $\chi_0(X)$ and is given by:
 \begin{align}
  \mvec B_\pm=\int \mrm d\mvec\rho\ \left[\mvec B_\mrm{ext}+\mvec B_\mrm M (\mvec\rho)\right]\left|\phi_\pm(\mvec\rho)\right|^2.
	\label{eq:eff-field}
 \end{align}
To approximate the average magnetic fields $\mvec B_\mrm{L,R}$ associated with the localized orbitals, in our simulations we take
\begin{align}
  \label{eq:local_fieldsL}
  \mvec B_\mrm L&\equiv\mvec B_- (\veps=-\veps_\mrm{max}),\\
	\label{eq:local_fieldsR}
  \mvec B_\mrm R&\equiv\mvec B_- (\veps=+\veps_\mrm{max}),
\end{align}
where $\ve_\mrm{max}$ is the largest positive detuning at which $\phi_\pm(\mvec \rho)$ are numerically calculated (for simulations in this paper, we choose $\ve_\mrm{max}=200\,\mu e\mathrm{V}$, which is larger than the typical tunnel splitting we consider, $\Omega\lesssim 100\,\mu e\mathrm{V}$, guaranteeing a localized state). Equations \eqref{eq:eff-field}, \eqref{eq:local_fieldsL}, and \eqref{eq:local_fieldsR} can then be used to estimate the parameters in Eq.~\eqref{eqnHlocal}.  Specifically, in our simulations we take $b=|g^*\mu_\mathrm{B}\mvec B|=|g^*\mu_\mathrm{B}\mvec B_-(\varepsilon=0)|$ [for strongly localized single-dot orbitals, $\mvec B_-(\varepsilon=0) \simeq (\mvec B_\mrm L+\mvec B_\mrm R)/2$] and we take $\Delta\mathbf{b}=g^*\mu_B\Delta\mvec B=g^*\mu_B(\mvec B_\mrm L-\mvec B_\mrm R)$, with $\mvec B_{\mrm{L},\mrm{R}}$ given by Eqs.~\eqref{eq:local_fieldsL} and \eqref{eq:local_fieldsR}.

We now evaluate $\mvec B$ and $\D\mvec B$ within the system of coordinates $(x,y,z)$ introduced in Sec.~\ref{secInteractions}. In this system of coordinates, we take the $z$-axis to be defined by the spin quantization axis at zero detuning ($\ve=0$). Figure~\ref{device_and_simulations}(c) shows $\Delta B^{x,z}$ as a function of the displacement $d_\mathrm{h}$ locating the nanomagnet for typical GaAs double-dot parameters (see Fig.~\ref{device_and_simulations} caption). The specific double-dot parameters chosen here were selected to achieve a tunnel splitting $\W/h=\W_\mathrm{opt}/h=19.1$~GHz, optimizing the average state-transfer fidelity in the presence of spin relaxation (see Sec.~\ref{secRelaxation}, below). For a symmetric configuration ($d_\mathrm{h}=0$), $\Delta B^z=0$ while $\Delta B^x\ne 0$ [left inset of Fig.~\ref{device_and_simulations}(c)], leading to a purely transverse spin-charge coupling. Notably, there is an asymmetric configuration for which $\Delta B^x$ vanishes while $\Delta B^z$ reaches a maximum [right inset of Fig.~\ref{device_and_simulations}(c)], leading to a purely longitudinal coupling. The type of spin-charge coupling can therefore be controlled simply through the relative position of the nanomagnet and the double quantum dot. While the position of the nanomagnet cannot be tuned \textit{in situ}, the single electron spin could be moved in a linear array of quantum dots~\cite{baart2016single}. For a nanomagnet position $d_\mathrm{h}$ such that $\Delta B^{z(x)}$ vanishes, the corresponding non-vanishing fields $\Delta B^{x(z)}$ reach 0.18~T with the parameters chosen here.

As described in Sec.~\ref{secInteractions}, electric-dipole coupling of the double dot to a superconducting coplanar waveguide resonator can be achieved by connecting one of the electrostatic gates of the double quantum dot to the central conductor of the resonator~\cite{childress2004mesoscopic,frey2012dipole}. In Fig.~\ref{device_and_simulations}(a), an unbiased gate located above the right dot is used as the coupling gate with a lever arm estimated to be $\alpha\simeq 0.2$~\cite{frey2012dipole}. For a microwave resonator with an impedance $Z_0=50\,\Omega$, a frequency $\wrs/2\pi=3\,\mathrm{GHz}$ and a zero-point voltage $\Vrms=\hbar\wrs\sqrt{2Z_0/h}=0.77\,\mu\mathrm{V}$, we estimate a charge-resonator coupling $e\alpha\Vrms/2h\simeq19\,\mathrm{MHz}$, in good agreement with previous experiments~\cite{frey2012dipole,toida2013vacuum,basset2013single,basset2014evaluating,stockklauser2015microwave}. For fields $\Delta B^{x,z}\simeq 0.18\,\mathrm{T}$ (estimated above) and for $g^*=-0.44$ in GaAs, the spin-charge coupling strengths are $\Delta b^{x,z}/4h\simeq 280\,\mathrm{MHz}$.  For these parameters and an estimated tunnel splitting  $\W/h=19.1\,\mathrm{GHz}$, the conditions for the perturbative effective Hamiltonian given following Eq.~\eqref{eqngz} are very well satisfied.

\subsection{Transverse spin-resonator coupling \label{secTransverse}}

\begin{figure}
\centering
\includegraphics*[width = 1.00\columnwidth]{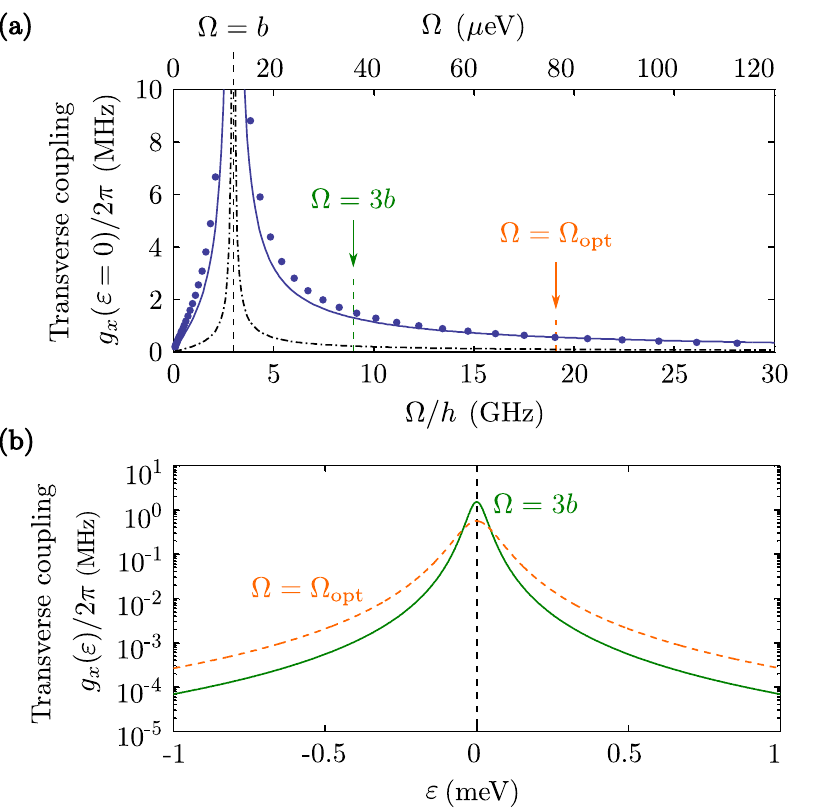}
\caption{Transverse spin-resonator coupling and electrical control. (a) Transverse coupling $g_x/2\pi$ from Eq.~\eqref{eqngx} at zero detuning ($\ve=0$) and on resonance ($b/h=\wrs/2\pi=3$~GHz) as a function of the tunnel splitting $\W$ for a constant $\D B^{x}=0.179$~T (blue line) and constant $\D B^{x}=0.030$~T (black dot-dashed line) for $e\alpha\Vrms/2h=19~\mathrm{MHz}$.  Blue circles account for a small change in $\D B^{x}$ as the tunnel-splitting is varied by tuning the double-dot separation (see the discussion at the start of Sec.~\ref{secTransverse} for details).  An optimal tunnel splitting $\Omega/h=\Omega_\mathrm{opt}/h=19.1$~GHz is indicated for $\D B^{x}=0.179$~T.  Selecting $\Omega=\Omega_\mathrm{opt}$ minimizes the effects of resonator damping and spin relaxation in the presence of electron-phonon coupling for the typical GaAs device parameters selected here (see Sec.~\ref{secRelaxation} and Fig.~\ref{figRelax}, below). (b) Transverse coupling $g_x/2\pi$ as a function of the detuning $\ve$ for a tunnel splitting of $\W/h=3b/h=9$~GHz (green line) and for the optimal tunnel splitting $\W/h=\W_\mathrm{opt}/h$ (orange dot-dashed line) of Fig.~\ref{figRelax}.}
\label{effectivecoupling}
\end{figure}

We now focus on the transverse spin-resonator coupling, enabling quantum state transfer between the spin and the resonator. To this end, we set the two systems on resonance at $b/h=\wrs/2\pi=3$~GHz and we choose $d_\mathrm{h}=0$ so that the longitudinal coupling is vanishingly small ($g_z\simeq 0$). As illustrated in Fig.~\ref{figDevice}(d), the spin-resonator coupling is mediated by the orbital degree of freedom, leading to an $\varepsilon$- and $\Omega$-dependent transverse coupling $g_x$. Figure~\ref{effectivecoupling}(a) shows $g_x$ at $\varepsilon=0$ evaluated from  Eq.~\eqref{eqngx} as a function of $\Omega$ with a fixed magnetic field distribution estimated in Fig. \ref{device_and_simulations}(c), $\Delta B^x=0.179\,\mathrm{T}$ (blue solid line) and, for comparison, with a fixed value $\Delta B^x=0.030\,\mathrm{T}$ (black dash-dotted line).  In addition, we have calculated $g_x$ by numerically solving the Schr\"odinger equation for the double-dot potential given in Eq.~\eqref{eq:quartic} for $\varepsilon =0$ and a range of double-dot separations $2a$ (taking $a=69\,\mathrm{nm}$ to $a=118\,\mathrm{nm}$ in steps of $1\,\mathrm{nm}$).  The resulting tunnel splitting $\epsilon_\mathrm{d}=\Omega$ (at $\varepsilon=0$) and the double-dot orbitals $\phi_\pm(\mvec\rho)$ are then used to determine $\D B^x=B^x_\mathrm{L}-B^x_\mathrm{R}$ via Eqs.~\eqref{eq:local_fieldsL} and \eqref{eq:local_fieldsR}.  Substituting these values into Eq.~\eqref{eqngx} for $g_x$ results in the blue dots in Fig.~\ref{effectivecoupling}(a), which account for a weak dependence of $\D B^x$ on $\Omega$ that we expect to be present in a real device. 

The all-resonant situation $\W=b=\hbar\wrs$ leads to a divergent coupling strength as shown in Fig.~\ref{effectivecoupling}(a). This is expected from the perturbation theory described in Sec.~\ref{secInteractions}, which breaks down for $\Omega=b$ ($\epsilon_\mathrm{d}=\sqrt{\varepsilon^2+\Omega^2}=\Omega$ when $\varepsilon=0$) and for $b=\hbar\omega_\mrm r$. If instead we choose a tunnel splitting $\W/h=3b/h=9$~GHz [thus respecting the criteria for perturbation theory given following Eq.~\eqref{eqngz}], we find $g_x/2\pi = 1.5\,\mathrm{MHz}$. This is more than four orders of magnitude stronger than the magnetic-dipole coupling between the spin and the resonator ($g_m/2\pi\sim 100\,\mathrm{Hz}$~\cite{imamoglu2009cavity}).

The transverse coupling is maximized when the electron is delocalized in both dots at $\ve=0$. As shown in Fig.~\ref{effectivecoupling}(b), moving away from this zero-detuning point strongly suppresses the coupling. The on/off ratio of the transverse coupling reaches more than $10^3$ for a reasonable detuning of 1~meV. Changes of the detuning on a time scale much faster than $1/g_x$ can be realized using the detuning gates identified in Fig.~\ref{device_and_simulations}(a)~\cite{petta2005coherent}, therefore realizing a tunable spin-resonator coupling.

\section{Sources of error	\label{secCharge}}

As shown in Sec.~\ref{secInteractions}, the spin-resonator coupling is mediated by the orbital degree-of-freedom. This coupling may lead to dephasing and relaxation in the presence of charge noise and the electron-phonon interaction. Here, we characterize the error in a quantum state transfer between the qubit and resonator, resulting from these error sources.  We follow a similar reasoning and notation to that given in Ref.~\cite{beaudoin2016hamiltonian} for analogous error sources. We quantify errors via the average fidelity
\begin{align}
 F=\int d\y \bra\y U_0^\dag\mathcal M(\proj\y)U_0\ket\y.	\label{eqnDefF}
\end{align}
In Eq.~\eq{eqnDefF}, $\ket\y=\cos(\vtt/2){\ket{\dwna0}}+\eul{i\phi}\sin(\vtt/2){\ket{\upa0}}$ is an arbitrary initial pure qubit state. Here, we take the qubit to be initialized in the dressed orbital groundspace, $\ket{s n}=\ket{s-n}'=\eul{-S}\ket{s-n}$, introduced in Sec.~\ref{secInteractions}. In the case of a quantum state transfer, $U_0$ is the ideal (error-free) unitary, $U_0\ket\y=\cos(\vtt/2){\ket{\dwna0}}-i\eul{i\phi}\sin(\vtt/2){\ket{\dwna1}}$, where the $i$ phase factor appears because the state transfer described here is equivalent to an SU(2) rotation. $\mathcal M$ is a completely-positive trace-preserving map representing the actual state transfer, including error from, e.g., cavity damping, spin relaxation, and dephasing. Finally, the integral represents an average with respect to the Haar measure $d\psi$, here equivalent to a uniform average over pure states on the surface of the Bloch sphere.  Errors are characterized by a finite infidelity, $1-F\ne 0$.

\subsection{Low-frequency charge noise	\label{secDephasage}}

Low-frequency charge noise can be an important source of dephasing for spin qubits whenever the qubit energy splitting depends on the double-dot detuning $\ve$~\cite{dial2013charge,harvey-collard2015nuclear}. Here, we account for this dephasing mechanism through a stochastic double-dot detuning term $V_\mrm{d}=\xi\tau_z/2$, leading to the total Hamiltonian $H_\mrm{tot}=H+V_\mrm d$, with $H$ given in Eq.~\eqref{eqnHlocal}. We take $\xi$ to be a Gaussian random variable with expectation value $\mrm E(\xi)=0$ and variance $\mrm E(\xi^2)=\sigma_\xi^2$.  As in Sec.~\ref{secInteractions}, we derive an approximate effective low-energy Hamiltonian, projected onto the dressed orbital groundspace (see \ref{appVdDerivation}): $H'_\mrm{tot}=H'+V'_\mrm{d}$.  Choosing $\varepsilon=0$ to maximize the spin-resonator coupling gives an effective spin-dephasing term
\begin{align}
  V_\mrm d' =  -\frac{\Delta b^z}{4\W} \xi\s_z,\quad (\varepsilon =0).	\label{eqnVprimed}
\end{align}
From Eq.~\eq{eqnDefF}, we then evaluate the fidelity of a quantum state transfer, accounting for inhomogeneous broadening in $\xi$ through the map $\mathcal M(\proj\y)=\mrm E(\eul{-i H'_\mrm{tot} t/\hbar}\proj\y\eul{i H'_\mrm{tot} t/\hbar})$, taking $b=\hbar\omega_r$ and $\varepsilon =0$. The state transfer is complete (minimizing the leading-order error) at $t=\pi/2g_x$. This gives the infidelity (error) to second order in the dimensionless small parameter $\D b^z\s_\xi/(\W \hbar g_x)\ll 1$:
\begin{align}
  1-F\simeq \frac{12+\pi^2}{24}\left(1-\frac{b^2}{\W^2}\right)^2\left(\frac{\D b^z}{\D b^x}\right)^2\left(\frac{\s_\xi}{e\al \Vrms}\right)^2,	\label{eqnFdeph}
 \end{align}
where we have used the expression for $g_x$ given in Eq.~\eq{eqngx}.

The error described by Eq.~\eq{eqnFdeph} can be substantial in a realistic setting. According to Fig.~\ref{device_and_simulations}(c), $\D b^z$ can be suppressed by proper placement of the nanomagnet above the double quantum dot. However, the magnetic-field simulations presented in Sec.~\ref{secIndiv} imply that a nanomagnet misalignment of $10\;\mrm{nm}$ would lead to $\D b^z/\D b^x\simeq0.08$. Fluctuations of the double-dot detuning $\s_\xi\simeq10\;\upmu\mrm{eV}$ have been reported in both GaAs~\cite{dial2013charge} and silicon~\cite{harvey-collard2015nuclear} due to charge noise, leading to $\s_\xi/e\al\Vrms\simeq50$ for $e\al\Vrms\simeq0.2\;\upmu\mrm{eV}$ (Sec.~\ref{secIndiv}). For such strong detuning noise, and assuming $b/\W\ll1$, Eq.~\eq{eqnFdeph} leads to an error of order 1. In the event of such a misalignment, for a very slowly-varying noise source, the error can be suppressed by narrowing the distribution of possible values (learning the value of $\xi$) through parameter estimation~\cite{klauser2006nuclear,sergeevich2011characterization,shulman2014suppressing} or by performing a dynamical decoupling sequence that averages the contribution $\sim \xi$ to zero~\cite{viola1999dynamical}. In particular, error due to inhomogeneous broadening can be suppressed by combining the Carr-Purcell dynamical-decoupling sequence (a train of $\pi$-pulses) with square-wave modulation of $g_x(t)$ in the SQUADD (SQUare wave And Dynamical Decoupling) protocol~\cite{beaudoin2016hamiltonian}. The error due to inhomogeneous broadening under SQUADD is given directly from Eq.~(4) of Ref.~\cite{beaudoin2016hamiltonian}. Inserting a noise amplitude $\D b^z\s_\xi/(2\W)$ [following Eq.~\eq{eqnVprimed} of this paper] and using the expression for $g_x$ given by Eq.~\eq{eqngx} in this paper, from Eq.~(4) of Ref.~\cite{beaudoin2016hamiltonian} we find a small error, $1-F\simeq0.8\%$, for a moderate number of decoupling $\pi$ pulses, $n_p=30$. For larger $n_p$ the error is rapidly suppressed ($\propto1/n_p^4$). Importantly, SQUADD suppresses any low-frequency noise afflicting the spin qubit, including nuclear-spin noise. This sequence is thus especially relevant for GaAs, since all nuclear isotopes of Ga and As carry spin.

\subsection{Orbital relaxation	\label{secRelaxation}}

As mentioned above, to describe state transfer between the spin and the resonator, we derive a projected effective Hamiltonian $H'$ acting on the manifold of low-energy dressed states $\left\{\ket{s-n}'\right\}$. Spin-charge and charge-resonator couplings result in an admixture of the bare excited double-dot state $\ket+$ in the dressed ground space of the double dot. This provides a mechanism through which the spin and resonator can relax into phonons via electron-phonon coupling~\cite{hu2012strong}.

To evaluate the state-transfer error that results from this relaxation mechanism, we derive a low-energy effective projected Hamiltonian starting from the electron-phonon interaction (see \ref{appEffectivePhonon}).  In this section, we assume $\Delta b^z\simeq 0$ so that pure-dephasing due to charge noise, phonons, and the longitudinal-coupling term are all negligible.  In particular, we assume the dominant sources of error are cavity damping, spin flips and the conversion of resonator photons into phonons, as described above.  In this limit, we derive a master equation for the system using a standard Born-Markov treatment of dissipation~\cite{gardiner2000quantum}. We take $\ve=0$ to maximize $g_x$, assuming $\W>b$ and $\W>\hbar\wrs$. We further neglect phonon absorption at low temperature, $\kB T\ll\min(\W-b,\W-\hbar\wrs,b,\hbar\wrs)$. In addition, we take the spin and the resonator to be resonant (neglecting small energy shifts): $b=\hbar\wrs$. This leads to the effective master equation for the density operator $\rho(t)$ in the low-energy subspace of spin-resonator states ($\left\{\ket{sn}=\ket{s-n}^\prime\right\}$): 
\begin{align}
  \dot\rho(t)=\liouv\rho(t)&=-i[H',\rho(t)]/\hbar+\kappa_0\dissip[a]\rho(t)\notag\\
  &\qquad+\g_\mrm d(b)\dissip[u_\mrm{sd}\s_--u_\mrm{dr} a]\rho(t),\label{eqnMaitresse}
\end{align}
where $\kappa_0$ is the bare resonator decay rate in the absence of the double dot, the damping superoperators are given by $\dissip[O]\rho(t)=O\rho(t) O^\dag-\frac12[O^\dag O\rho(t)+\rho(t)O^\dag O]$ for a general operator $O$, $\gamma_\mathrm{d}(\Delta\epsilon)$ gives the orbital decay rate due to phonons for the double-dot with orbital energy splitting $\Delta\epsilon$, and we have introduced the dimensionless parameters (see also \ref{appEffectivePhonon} with $\varepsilon =0$, $b=\hbar\omega_\mathrm{r}$):
\begin{align}
  u_\mrm{sd}&=\frac{\W\,\D b^x/2}{\W^2-b^2},
 \hspace{10mm}
 u_\mrm{dr}=\frac{\W\,e\al\Vrms}{\W^2-b^2}.	\label{eqnu}
\end{align}
The parameters $u_\mathrm{sd}$ and $u_\mathrm{dr}$ quantify the degree of admixture of $\ket +$ in the subspace $\{\ket{s-n}'\}$. The perturbative approach presented here will yield accurate predictions when $u_\mrm{sd}\ll1$, $u_\mrm{dr}\ll1$.  In the second line of Eq.~\eqref{eqnMaitresse}, operators describing spin and resonator decay appear within the same dissipator, since these processes can interfere for $b\simeq\hbar\w_\mrm r$.  

In GaAs, both deformation-potential and piezoelectric coupling mechanisms are relevant and the conduction band has a non-degenerate minimum.  The rate $\gamma_\mathrm{d}(b)$ can then be calculated explicitly in the long-wavelength limit giving (from, e.g., taking the limit $\kB T\ll b$ in Eq.~(56) of Ref.~\cite{beaudoin2015microscopic}),
\begin{align}
 \g_\mrm d(b)&=\left[\frac{\Xi^2}3\!\left(\!\frac {b}{\hbar v_\mrm{LA}}\!\right)^4\!\!
		  +\frac4{35}\!\left(\!1+\frac{4\z^2}3\right)\!\left(\frac{e\,e_{14}}{\ve_\mrm{diel.}}\frac{b}{\hbar v_\mrm{LA}}\right)^2\right]\notag\\
	&\hspace{1cm}\times\frac{9\pi}{\hbar^2}\frac{b}{m_\mrm{at}\w_\mrm D^3}|\mvec\sq|^2.
    \label{eqngammad}
\end{align}
For GaAs, $\Xi=-8.6$~eV is the volume deformation potential for the conduction band minimum, $\z=v_\mrm{LA}/v_\mrm{TA}$, $v_\mrm{LA}=5210\;\mrm{m/s}$ and $v_\mrm{TA}=3070\;\mrm{m/s}$ are the phase velocities of the transverse and longitudinal acoustic branches, respectively, $e$ is the fundamental electric charge, $e_{14}=-0.16\;\mrm{C/m^2}$ is the only nonvanishing element of the piezoelectric tensor, $\ve_\mrm{diel.}=12.9\,\ve_0$ is the static dielectric constant, $m_\mrm{at}=1.20\times10^{-25}\;\mrm{kg}$ is the mass per lattice atom, and $\w_\mrm D/2\pi=7.50\;\mrm{THz}$ is the Debye frequency~\cite{ioffe1998}. Finally, in Eq.~\eq{eqngammad},  we have introduced the transition dipole matrix element of the double quantum dot, $\mvec\sq=\int d\mvec\rho\,\mvec\rho\,\phi^\ast_+(\mvec\rho)\phi_-(\mvec\rho)$. Equation~\eq{eqngammad} relies on a long-wavelength approximation (see \ref{appEffectivePhonon}), which will lead to accurate predictions when the double-dot separation $2a$ is much smaller than the minimum wavelength of emitted phonons, $\ld=h v_\mrm{TA}/b$. For $b/h=\omega_\mathrm{r}/2\pi=3$~GHz, this condition ($2a\ll h v_\mrm{TA}/b$) translates to $2a\ll1$~$\upmu$m in GaAs.

\begin{figure}
 \begin{center}
  \includegraphics[width=0.48\textwidth]{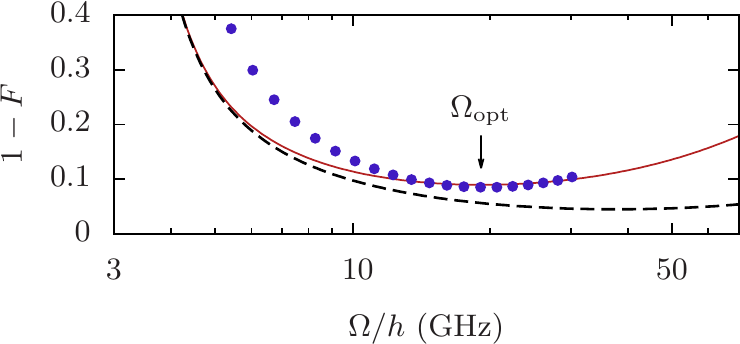}
 \end{center}
 \caption{State-transfer error due to spin and cavity decay in the presence of electron-phonon coupling and cavity damping [Eq.~\eq{eqnRelax}].  The error shows a minimum at the optimal tunnel splitting $\W=\W_\mathrm{opt}$. The parameters used for this plot are: $\wrs/2\pi=b/h=3$~GHz, $\D b^x/4h=294$~MHz, $e\al\Vrms/2h=18.7$~MHz, $|\mvec\sq|=67.8$~nm. Material parameters for GaAs are given in the main text following Eq.~\eqref{eqngammad}. Solid red line: $\kp/2\pi=0.1$~MHz. Dashed black line: $\kp/2\pi=0.025$~MHz. Blue dots: error calculated with Eq.~\eq{eqnRelax}, accounting for the dependence of $\D b^x$ and $|\mvec\sq|$ on the double-dot separation (which controls $\W$).  The blue dots in this figure were produced using the same procedure described at the beginning of Sec.~\ref{secTransverse}. \label{figRelax}}
\end{figure}

To evaluate the fidelity, we substitute $\mathcal M[\proj\y]= \eul{\liouv t}\proj\y$ into Eq.~\eq{eqnDefF}, with $\liouv$ the Lindbladian defined in Eq.~\eq{eqnMaitresse}. We assume that the error is small (i.e., that the state-transfer time $t=\pi/2g_x$ is short compared to the typical spin-relaxation and resonator damping times).  To calculate the fidelity in this limit, we expand the propagator $\eul{\liouv t}$ for small $t=\pi/2g_x$, collecting terms at leading order in the dimensionless small parameters $\kp_0/g_x$, $u_\mrm{sd}^2\g_\mrm d(b)/g_x$, and $u_\mrm{dr}^2\g_\mrm d(b)/g_x$. Substituting the expressions for $g_x$, $u_\mrm{sd}$, and $u_\mrm{dr}$ given in Eqs.~\eq{eqngx} and \eq{eqnu}, we find an infidelity
\begin{align}
  1-F &\simeq \frac\pi3\left[\frac{|\W^2-b^2|}{e\al\Vrms\,\D b^x}\frac{\hbar\kappa_0}{\W}\right.\notag\\
  &\;+\left.\frac{(e\al \Vrms)^2+(\D b^x/2)^2}{e\al\Vrms\,\D b^x}\frac{\W\,\hbar\g_\mrm d(b)}{|\W^2-b^2|}\right].
  \label{eqnRelax}
 \end{align}
The first term in Eq.~\eq{eqnRelax} gives the error arising from intrinsic resonator damping ($\propto\kappa_0$). The second term [$\propto\gamma_\mathrm{d}(b)$] describes the error from decay of the spin and resonator into phonons. Figure~\ref{figRelax} shows the error given by Eq.~\eq{eqnRelax} as a function of $\W/h$ for $\kappa_0/2\pi=0.1$~MHz (solid red line) and $\kappa_0/2\pi=0.025$~MHz (dashed black line). 

The infidelity given by Eq.~\eqref{eqnRelax} diverges at a ``hot spot'' for $\W=b$ due to enhanced decay into phonons arising from an increased admixture of the bare double-dot excited state~\cite{stano2006orbital,stano2006theory,hu2012strong}.  Designing a double-dot with a large tunnel splitting $\W\gg b$ to avoid the hot spot suppresses phonon emission, but will also increase the required transfer time ($t\propto 1/g_x\propto \Omega$ in this limit), leading to a larger error due to cavity damping. Because of this competition, there is an optimal value of $\W>b$ that minimizes the error. The value $\W=\W_\mathrm{opt}$ that minimizes error in Eq.~\eq{eqnRelax} is
\begin{align}
 \W_\mathrm{opt}&=b\;\sqrt{\frac{1+r+\sqrt{1+2r}}r},	\label{eqnOmegaOpt}\\
 r&=\frac{2\kp_0}{\g_\mrm d(b)}\frac{b^2}{(e\al\Vrms)^2+(\D b^x/2)^2}.	\label{eqnr}
\end{align}

The optimal (maximum) fidelity $F=F_\mathrm{max}$ for $\W=\W_\mathrm{opt}$ is given directly by substituting Eq.~\eq{eqnOmegaOpt} into Eq.~\eq{eqnRelax}. This gives
\begin{align}
 1-F_\mrm{max}&\simeq\frac\pi3\sqrt{\frac{\kp_0 \g_\mrm d(b)}{\tilde{g}^2}},		\label{eqnFmax}
\end{align}
where $\hbar\tilde{g}$ is an energy scale that interpolates between the spin-charge coupling ($\sim \D b^x/2$) and the charge-resonator coupling ($\sim e\alpha V_\mathrm{rms}^0$):
\begin{align}
 \hbar \tilde{g}&=\frac{e\al\Vrms\;\D b^x/2}{\sqrt{(e \al \Vrms)^2+(\D b^x/2)^2}}.		\label{eqngeff}
\end{align}

For the typical Zeeman splitting considered here ($b/h\lesssim 3$ GHz) and in a GaAs device, the phonon-induced decay rate $\g_\mrm d(b)$ given by Eq.~\eq{eqngammad} is dominated by the piezoelectric (rather than deformation-potential) mechanism.  For piezoelectric phonons [the contribution $\propto e_\mathrm{14}^2$ in Eq.~\eqref{eqngammad}], $\g_\mrm d(b)\propto b^3|\mvec\sq|^2$. In addition, for a typical achievable magnetic field gradient of $\D B^x\simeq 0.18\,\mathrm{T}$ in GaAs ($g^*=-0.44$), and for $e\al \Vrms\simeq 0.2\,\mu e\mathrm{V}$ (estimated above), we have $|\D b^x|/2\gg e\al \Vrms$.  In this limit, $\tilde{g}\simeq e\al\Vrms\propto \hbar\omega_\mathrm{r}$, where we have used $\Vrms=\hbar\omega_\mathrm{r}\sqrt{2 Z_0/h} \propto\hbar\wrs$ at fixed impedance $Z_0$. Substituting the above scaling relations into Eq.~\eq{eqnFmax} and taking $\hbar\omega_\mathrm{r}=b$ on resonance then leads to $1-F_\mrm{max}\propto|\mvec\sq|\sqrt{\kp_0 \omega_\mathrm{r}}$. To suppress the error at the optimal tunnel splitting, it is thus necessary to minimize not only cavity damping, but also the resonator frequency and the dipole moment of the double quantum dot.

In Eqs.~\eq{eqnOmegaOpt} to \eq{eqnFmax}, as a first approximation, we have assumed that $\D b^x$ and $|\mvec\sq|$ are independent of $\W$. In practice, however, $\W$ is tuned by changing the distance between the two dots, leading to a weak correlation between $\D b^x$, $|\mvec\sq|$ and $\W$. We have accounted for this correlation by numerically solving the Schr\"odinger equation for a range of double-dot separations [similar to the procedure described in Sec.~\ref{secTransverse} to determine the blue dots in Fig. \ref{effectivecoupling}(a)].  This procedure gives $\D b^x$ and $|\mvec\sq|$ [which can be evaluated from the double-dot orbitals $\phi_\pm(\mvec \rho)$] as a function of $\W$ (given by the ground-state orbital level splitting at $\ve=0$). Substituting the results into Eq.~\eq{eqnRelax} then gives the blue dots shown in Fig.~\ref{figRelax}.

Finally, in this section, we have neglected contributions to spin relaxation arising from intrinsic spin-orbit coupling~\cite{khaetskii1999spin,golovach2004phonon}. Theoretical treatments of spin-orbit coupling have led to the prediction of hot spots in the spin relaxation rate similar to that shown in Fig.~\ref{figRelax}~\cite{stano2006orbital,stano2006theory}. However, we expect the decay rate due to spin-orbit coupling to be negligible compared to the decay rates obtained above. Indeed, Fig.~4 of Ref.~\cite{stano2006orbital} gives a spin-relaxation rate $\simeq10^2$--$10^3\;\mrm s^{-1}$ for a single spin in a double quantum dot when taking $\W/h=19.1$~GHz, $B=0.6$~T, and a single-dot Bohr radius (confinement length) $\simeq 30$~nm. In contrast, Eq.~\eq{eqnMaitresse} gives a much larger rate $\simeq10^5\;\mrm s^{-1}$, also assuming $\W/h=19.1$~GHz and using the parameters given in the caption of Fig.~\ref{figRelax}, above.

\section{Discussion}\label{secDiscussion}

The fidelity of the quantum state transfer is optimized in the previous section by fixing the Zeeman splitting $b$ to a close-to-minimal value of 3~GHz. The minimization of thermal excitations with $\mrm{min}(\eps_\mrm d,b,\hbar\wrs)\gg k_\mrm BT$ sets a lower bound on $b=\hbar\wrs$ to approximately 1~GHz for $T=50$~mK. A second constraint is set by the minimal external field $B_\mrm{ext}$ necessary to fully magnetize the nanomagnet. For the design considered here, the external magnetic field at saturation is approximately $0.6$~T~\cite{lachance2015magnetometry}. This puts a lower bound on $b/h$ to 2.9~GHz, which could be reduced by increasing the shape anisotropy of the nanomagnet~\cite{lachance2015magnetometry,forster2015electric}.

In the proposed device, obtaining a 90\% fidelity for a spin-resonator state transfer requires a resonator with a linewidth of 0.1~MHz. Linewidths of superconducting resonators on GaAs are currently limited to approximately $0.4$~MHz, probably due to the piezoelectricity of GaAs~\cite{toida2013vacuum,frey2013interaction}. Etching away the substrate in the gap between the central conductor and the ground plane should therefore reduce the linewidth~\cite{barends2010reduced,bruno2015reducing}. Losses from vortices induced by an external field can be limited using superconducting resonators designed to minimize vortex formation. In such resonators, linewidths below 0.02~MHz in an in-plane field up to 6~T have recently been demonstrated~\cite{samkharadze2016high}. A resonator linewidth below 0.1~MHz seems therefore feasible through careful engineering of superconducting resonators on GaAs.

With all these ingredients, a proof-of-principle experiment demonstrating coherent coupling of a single spin to a microwave resonator may be possible in GaAs. Furthermore, the proposed coupling scheme could be implemented in other materials as it does not rely on intrinsic properties of the host semiconductor of the spin qubit. The absence of nuclear spins and piezoelectric interaction with phonons in silicon should result in much better fidelities. Like in GaAs, fast electrical control of the spin qubit~\cite{yoneda2014fast,takeda2016a} could be used to mitigate charge dephasing using SQUADD~\cite{beaudoin2016hamiltonian}.

In conclusion, we have considered a single-electron spin in an inhomogeneous magnetic field in a double quantum dot coupled to a microwave resonator. We have shown that both transverse and longitudinal interactions between the spin and the resonator arise in this system. While the transverse interaction is useful for a quantum state transfer between the spin and the resonator, the longitudinal interaction could be used to perform quantum nondemolition spin readout~\cite{didier2015fast} and two-qubit gates between distant spins~\cite{jin2012strong,royer2016fast}. We have also shown that with proper placement of a nanomagnet in a quantum-dot device, either purely transverse or purely longitudinal spin-charge couplings can be achieved, leading to a spin-resonator interaction strength reaching a few MHz. Spin dephasing and relaxation, enhanced through the spin-charge interaction, have been investigated in the context of a state transfer. Using dynamical decoupling and careful parameter optimization, we predict that a state-transfer fidelity above 90\% in a GaAs device is within experimental reach.

\section*{Acknowledgments}
We thank Alexandre Blais and Julien Camirand Lemyre for useful discussions. This work was supported by NSERC, CIFAR, FRQNT, CFI and the W. C. Sumner Foundation.

\appendix

\section{Effective Hamiltonians \label{appSW}}

In this Appendix, we derive the effective coupling Hamiltonian $H'$ [Eq.~\eqref{eqnHprime}], the effective pure-dephasing Hamiltonian due to charge noise, $V_\mathrm{d}'$ [Eq.~\eqref{eqnVprimed}], and the effective electron-phonon coupling Hamiltonian, $V_\mathrm{e-ph}'$, that gives rise to the master equation describing spin flips and dot-mediated resonator damping [Eq.~\eqref{eqnMaitresse}].

We first diagonalize the orbital part of the starting Hamiltonian $H$ [Eq.~\eqref{eqnHlocal}]:
\begin{align}
\tilde{H} =& R_y^\dagger(\theta) H R_y(\theta),\\
& R_y(\theta)=e^{-i\tau_y\theta/2},\quad\tan \theta = \frac{\Omega}{\varepsilon}.
\end{align}
The rotated Hamiltonian is then divided in two parts: $\tilde{H}=\tilde{H}_0+\tilde{V}$, where $\tilde{H}_0$ is diagonal and $\tilde{V}$ is purely off-diagonal in the basis $\{\ket{sdn}\}$ of simultaneous eigenstates of spin ($\sigma_z$), double-dot orbital ($\tau_z$), and microwave-photon number operator ($n=a^\dagger a$).  We then perform a further unitary (Schrieffer-Wolff) transformation:
\begin{multline}
e^S \tilde{H}e^{-S}=\tilde{H}_0+\tilde{V}+\left[S,\tilde{H}_0\right]\\
+\frac{1}{2}\left[S,\left[S,\tilde{H}_0\right]\right]+\left[S,\tilde{V}\right]+\ldots,\label{eqnSWTransformation}
\end{multline}
parametrized by the antiunitary operator $S^\dagger =-S$.  To approximately diagonalize $\tilde{H}$, we require that the terms $\sim \mathcal{O}(\tilde{V})$ in Eq.~\eqref{eqnSWTransformation} vanish by choosing an $S$ that satisfies
\begin{equation}
\tilde{V}+\left[S,\tilde{H}_0\right]=0.\label{eqnSWCondition}
\end{equation} 
The solution to Eq.~\eqref{eqnSWCondition} can be expressed formally as $S=\tilde{L}_0^{-1}\tilde{V}$ where $\tilde{L}_0O=[\tilde{H}_0,O]$.  Evaluating commutators term-by-term and inverting then gives 
\begin{align}
 S&=\frac{\ve \D b^x}{4b\eps_\mrm d}\tau_z\s_++\frac{\W\D b^z}{4\eps_\mrm d^2}\s_z\tau_-
    +\left(1-\frac{\ve\tau_z}{\eps_\mrm d}\right)\frac{e\al\Vrms}{2\hbar\wrs}a^\dag\notag\\
  &\qquad+\frac{\W\D b^x}{4\eps_\mrm d(\eps_\mrm d-b)}\,\s_+\tau_- 
    +\frac{\W\D b^x}{4\eps_\mrm d(\eps_\mrm d+b)}\,\s_-\tau_-\notag\\
  &\qquad+\frac{\W e\al\Vrms}{2\eps_\mrm d(\eps_\mrm d-\hbar\wrs)}\,a\tau_+
  +\frac{\W e\al\Vrms}{2\eps_\mrm d(\eps_\mrm d+\hbar\wrs)}\,a^\dag\tau_+\notag\\
  &\qquad\qquad-\hc.	\label{eqnS}
\end{align}
Finally, we project onto the low-energy orbital ground space (corresponding to $\tau_z=-1$ in $\tilde{H}$) with an asymmetric projection operator $P_-=\sum_{sn} \left|s-n\right>\left<sn\right|$ to arrive at an effective Hamiltonian for only the spin- and resonator degrees-of-freedom:
\begin{align}
H_\mathrm{eff}=& P_-^\dagger \left(e^S \tilde{H}e^{-S}\right) P_-\label{eqnHeff}\\
=&P_-^\dagger \left(\tilde{H}_0+\frac{1}{2}\left[S,\tilde{V}\right]\right) P_-+\mathcal{O}\left(\tilde{V}^3\right)\\
=&H'+H_\mathrm{CR}+H_\mrm{sq}+\mathrm{const.}+\mathcal{O}\left(\tilde{V}^3\right). 
\end{align}
The final spin-resonator Hamiltonian $H'$ given in Eq.~\eqref{eqnHprime} of the main text neglects counter-rotating terms $H_\mathrm{CR}\propto \Delta b^x\left(\sigma_- a+\mathrm{H.c.}\right)$, squeezing terms $H_\mrm{sq}\propto e\al\Vrms(a^2+a^{\dag 2})$, an overall constant, and further corrections of order $\sim\mathcal{O}(\tilde{V}^3)$. From Eq.~\eqref{eqnHeff}, it is clear that this procedure is equivalent to a direct projection of $H$ into the dressed orbital groundspace $\left\{\left|s-n\right>'\right\}=\left\{e^{-S}\left|s-n\right>\right\}$.

In addition to the spin-resonator coupling, the projected Hamiltonian $H'$ contains a Lamb shift for the spin,
\begin{equation}
 \chi_\mrm s = \frac{(\ve\D b^x)^2}{8b\eps_\mrm d^2}-\frac{b(\W\D b^x)^2}{8\eps_\mrm d^2(\eps_\mrm d^2-b^2)},
\end{equation}
and ac-Stark shifts for the spin ($\tilde{\chi}_\mrm s$) and the resonator ($\chi_\mrm r$):
\begin{align}
 \tilde\chi_\mrm s &= \frac{\veps\D b^z}{2\edt},	\hspace{5mm}
 \chi_\mrm r = \frac{(\W e\al\Vrms)^2}{2\eps_\mrm d(\eps_\mrm d^2-\hbar^2\wrs^2)}.
\end{align}

\subsection{Effective pure-dephasing Hamiltonian}\label{appVdDerivation}
To arrive at a leading-order effective Hamiltonian describing pure dephasing due to low-frequency charge noise, we follow a similar procedure to that outlined above.  Defining $\tilde{V}_\mathrm{d}=R_y^\dagger(\theta)V_\mathrm{d}R_y(\theta)$ with $V_\mathrm{d}=\xi \tau_z/2$, we find an effective dephasing Hamiltonian:
\begin{align}
V_\mathrm{d}^\mathrm{eff} =& P_-^\dagger\left(e^S \tilde{V}_\mathrm{d}e^{-S}\right)P_-,\\
=& P_-^\dagger\left(\tilde{V}_\mathrm{d}+\left[S,\tilde{V}_\mathrm{d}\right]\right)P_-+\mathcal{O}\left(\tilde{V}_\mathrm{d} \tilde{V}^2\right),\\
=& V_\mathrm{d}'+V_\mathrm{d,CR}+\mathcal{O}\left(\tilde{V}_\mathrm{d} \tilde{V}^2\right).
\end{align}  
The effective Hamiltonian $V_\mathrm{d}'\sim \mathcal{O}\left(V_\mathrm{d} \tilde{V}\right)$ given in Eq.~\eqref{eqnVprimed} of the main text for the special case $\varepsilon =0$ neglects counter-rotating terms $V_\mathrm{d,CR}$ ($\propto a,a^\dagger, \sigma_\pm$) and corrections that are of order $\sim \mathcal{O}\left(\tilde{V}_\mathrm{d} \tilde{V}^2\right)$ or higher. 

\subsection{Effective phonon coupling}\label{appEffectivePhonon}

Here we derive the effective spin-phonon and resonator-phonon couplings relevant for Sec.~\ref{secRelaxation} of the main text.  We begin from the general electron-phonon coupling (see, e.g., Ref.~\cite{mahan1990many})
\begin{equation}\label{eqnHe-ph}
H_\mathrm{e-ph} = \sum_{\mathbf{q}\lambda}A_{\mathbf{q}\lambda}\rho_\mathbf{q}(b_{\mathbf{q}\lambda}+b_{-\mathbf{q}\lambda}^\dagger),
\end{equation}
where $A_{\mathbf{q}\lambda}=A_{-\mathbf{q}\lambda}^*$ are coupling constants for phonons of wavevector $\mathbf{q}$ and branch $\lambda$ with associated annihilation operators $b_{\mathbf{q}\lambda}$.  The electron density operator can be written as $\rho_\mathbf{q}=\sum_{\alpha\beta}c_\alpha^\dagger c_\beta S_{\alpha\beta}(\mathbf{q})$ with form factors $S_{\alpha\beta}(\mathbf{q})=\int d^3 r e^{-i\mathbf{q}\cdot\mathbf{r}}\psi_\alpha^*(\mathbf{r})\psi_\beta(\mathbf{r})$.  Here, $c_\alpha$ annihilates an electron in orbital state $\psi_\alpha(\mathbf{r})$ ($\left\{\psi_\alpha(\mathbf{r})\right\}$ is taken to form a complete orthonormal set).  We project onto the double-dot orbital ground-state doublet $\left\{\ket{\pm}\right\}$ and make the long-wavelength approximation, $S_{\alpha\beta}(\mathbf{q})\simeq -i\mathbf{q}\cdot\bra{\alpha}\hat{\mathbf{r}}\ket{\beta}$.  For real wavefunctions $\psi_\pm (\mathbf{r})=\psi^*_\pm (\mathbf{r})$, the dipole transition matrix element is $\mvec\sq=\bra{+}\hat{\mathbf{r}}\ket{-}=\bra{-}\hat{\mathbf{r}}\ket{+}$.  At zero detuning ($\varepsilon =0$), the wavefunctions $\psi_\pm(\mathbf{r})$ are symmetric about $\mathbf{r}=0$ giving: $\bra{+}\hat{\mathbf{r}}\ket{+}=\bra{-}\hat{\mathbf{r}}\ket{-}=0$.  The projected density operator, written in the basis of states $\ket{\pm}$ then reads
\begin{equation}\label{eqnRhoqProjected}
\tilde{\rho}_\mathbf{q} = -i\mathbf{q}\cdot\mvec\sq\tau_x +\mathrm{const.}, 
\end{equation}
where here, $\tau_x=\ket{+}\bra{-}+\ket{-}\bra{+}$.  Substituting $\rho_\mathbf{q}$ in Eq.~\eqref{eqnHe-ph} with $\tilde{\rho}_\mathbf{q}$ and neglecting the constant term in Eq.~\eqref{eqnRhoqProjected} gives the effective electron-phonon coupling
\begin{equation}
\tilde{H}_\mathrm{e-ph} = X\tau_x;\quad X=\sum_{\mathbf{q}\lambda}\alpha_{\mathbf{q}\lambda}(b_{\mathbf{q}\lambda}+b_{-\mathbf{q}\lambda}^\dagger),
\end{equation}
where $\alpha_{\mathbf{q}\lambda}=-i\mathbf{q}\cdot\mvec\sq A_{\mathbf{q}\lambda}$. 

Following the same procedure described in \ref{appVdDerivation}, we then find the approximate effective spin-phonon and resonator-phonon interactions from the projected low-energy Hamiltonian:
\begin{eqnarray}
V^\prime_\mathrm{e-ph}&=&P_-^\dagger\left(\tilde{H}_\mathrm{e-ph}+\left[S,\tilde{H}_\mathrm{e-ph}\right]\right)P_-,\\
&=& \left[u_\phi\sigma_z+u_\mathrm{sd}\sigma_x-u_\mathrm{dr}(a+a^\dagger)\right]X,\label{eqnEffectivePhonons}
\end{eqnarray} 
where $u_\phi = \Omega\Delta b^z/(2\W^2)$, $u_\mathrm{sd}=\Omega(\Delta b^x/2)/(\W^2-b^2)$, and $u_\mathrm{dr}=\Omega e\alpha V^0_\mathrm{rms}/[\W^2-(\hbar\omega_\mathrm{r})^2]$.  The term $\sim u_\phi\propto \Delta b^z$ gives rise to spin dephasing in the presence of a phonon bath, the term due to the spin-dot coupling $\sim u_\mathrm{sd}\propto \Delta b^x$ gives rise to spin flips in combination with phonon absorption/emission, and the term due to the dot-resonator coupling $\sim u_\mathrm{dr}\propto V^0_\mathrm{rms}$ converts between microwave-resonator photons and phonons. 

The first term in Eq.~\eq{eqnEffectivePhonons} leads to loss of spin coherence \cite{palma1996quantum}. It may be that the error introduced by the term $\propto u_\phi$ is comparable to that induced by the terms $\propto u_\mathrm{sd},u_\mathrm{dr}$ at short times when $\Delta b^z$ is not negligible.  For simplicity, in the main text we have restricted our attention to the case of a near-ideal state transfer, where $\Delta b^z=0$ (and hence, $u_\phi =0$).


\section*{References}

\bibliography{article}
\bibliographystyle{iopart-num}

\end{document}